# Can periodicity in low altitude cloud cover be induced by cosmic ray variability in the extragalactic shock model?

**Dimitra Atri[1], Brian C. Thomas[2], and Adrian L. Melott[1]**

*1. Department of Physics and Astronomy, University of Kansas, Lawrence, Kansas 66045, USA*

*2. Department of Physics and Astronomy, Washburn University, Topeka, Kansas 66621, USA*

**ABSTRACT**

Variation in high energy cosmic rays (HECRs) has been proposed to explain a 62 My periodicity in terrestrial fossil biodiversity. It has been suggested that the infall of our galaxy toward the Virgo cluster could generate an extragalactic shock, accelerating charged particles and exposing the earth to a flux of high energy cosmic rays (HECRs). The oscillation of the Sun perpendicular to the galactic plane could induce 62 My periodicity in the HECR flux on the Earth, with a magnitude much higher than the Galactic cosmic ray change we see in a solar cycle. This mechanism could potentially explain the observed 62 My periodicity in terrestrial biodiversity over the past 500 My. In addition to direct effects on life from secondaries, HECRs induced air showers ionize the atmosphere leading to changes in atmospheric chemistry and microphysical processes that can lead to cloud formation including low altitude cloud cover. An increase in ionization changes the global electric circuit which could enhance the formation of cloud condensation nuclei (CCN) through microphysical processes such as electroscavenging and ion mediated nucleation, leading to an increase in the cloud cover. This could increase the albedo and reduce the solar flux reaching the ground, reducing the global temperature. Using an existing model, we have calculated the enhancement in atmospheric ionization at low altitudes resulting from exposure to HECRs. We use a conservative model to estimate the change in low altitude cloud cover from this increased ionization.

**INTRODUCTION**

A strong ~62 My periodicity in marine biodiversity has been observed in independent paleobiological databases, going back to 500 My (Rhode and Muller, 2005; Melott and Bambach, 2010 and references therein). There are two major hypotheses to explain this phenomenon. Certain features of the data suggest that uplift or sea level change may be involved, but this does not account for periodicity (Melott and Bambach, 2010). The extragalactic shock model hypothesis (Medvedev and Melott, 2007) can account for the periodicity by the infall of our galaxy toward the Virgo cluster coupled with the known solar oscillation perpendicular to the galactic plane (Gies and Helsel, 2005) with a period of about 62 My. The possible causal mechanisms, which need work, are based on enhancement of the HECR flux on the Earth (Medvedev and Melott, 2007). This enhanced flux results in an increase in the ionization rate in the atmosphere which leads to changes in the atmospheric chemistry (not unique to radiation events—see, e.g. Melott et al 2010a) These changes result in ozone depletion in the upper stratosphere, which increases the solar UVB flux on the ground, directly affecting the biosphere (Melott et al., 2010b). An ozone crisis can seriously impact life through the mechanism of a food-chain crash in the oceans (Melott and Thomas, 2009, and references therein.) HECRs will also result in an increased flux of high energy muons on the ground, which could directly affect living organisms. Muons also have the capability to penetrate deep below the land surface as well as through water. There will be an increase in thermal neutrons at ground level, which are also hazardous due to their very large cross sections. The magnitude of this damage is also under investigation. Based on the results of Juckett (2007, 2009) it should be far from negligible.

There is some evidence that cosmic ray induced atmospheric ionization could enhance the low altitude cloud cover (Svensmark and Friis-Christensen, 1997; Enghoff and Svensmark, 2008; Harrison et al., 2003; Harrison, 2000). Although the field of cloud microphysics is still in its infancy, preliminary experimentation suggests a correlation between the atmospheric ionization and cloud formation (Dulipssy et al., 2010). This ionization change will affect the global ionosphere-earth current density $J_z$, which has effects on cloud microphysical processes such as electroscavenging and ion mediated nucleation (Tinsley, 2008). Change in ionization from the cosmic ray variability has been previously computed for a solar cycle, which indicates a linear relationship between ionization and the low altitude cloud cover (Usoskin et al., 2004). However, other studies show a much lesser effect on the low altitude cloud cover. Most conservative estimates give between 10% (Kulmala et al., 2010) and 20% (Erlykin et al., 2009; Sloan and Wolfendale, 2008) contribution to the total change in cloud cover from galactic cosmic rays. We note that the cosmic rays in the extragalactic hypothesis (Medvedev and Melott, 2007) are of higher energy, and will penetrate deeper in the atmosphere (Atri et al., 2010) and should have a greater effect than the galactic cosmic rays. We therefore regard 10% as a lower limit to the correlation.

Muons are the primary source of atmospheric ionization at low altitudes. With an enhancement of HECR flux, there will be an increase in the flux of high energy muons at low altitudes, thereby enhancing atmospheric ionization there. This could lead to increased low altitude cloud cover resulting from HECR exposure, as predicted by the extragalactic shock model.

**METHOD**

The enhanced spectra from the extragalactic shock model results in an increase in the flux of primaries up to about 1 PeV. Because of the uncertainty in the values of the magnetic field parameters in this model, it gives a range of possible results. Here, we explore the maxima and minima of HECR exposure obtained from the model. As shown in Fig 1, the minimum flux (case 1) has the flux enhanced by a factor of 4 at a TeV and by a factor of 25 for the maximum flux (case 2). These primaries collide with the nuclei of $N_2$ and $O_2$ in the atmosphere, producing air showers. These consist of charged particles and photons which ionize the atmosphere, neutrons and other short-lived unstable particles. The component which ionizes the atmosphere is called the electromagnetic component of the shower and can break molecular bonds, leading to atmospheric chemistry changes. As the energy of protons hitting the atmosphere goes up, the total atmospheric ionization increases and so does the ionization in the lower atmosphere. We have used an existing model to compute atmospheric ionization as a function of altitude resulting from the HECR enhancement (Atri et al., 2010). This model provides the atmospheric ionization profiles from 10 GeV to 1 PeV primaries. Convolving the enhanced spectra from the extragalactic shock model with the cosmic ray ionization model thus gives us the corresponding maxima and minima of atmospheric ionization. The normal atmospheric ionization values corresponding to lower energies are calculated using an existing model (Usoskin and Kovaltsov, 2006). We concentrate on the increase in ionization from these two cases at an altitude of 3 km (~700 g cm$^{-2}$), where most low altitude clouds reside.

**RESULTS AND DISCUSSION**

We compared the values of atmospheric ionization from the two cases with that at the time of the cosmic ray maxima in a solar cycle. Looking at the total energy deposited in the atmosphere, case 1 has a global enhancement factor of 3.2 and case 2 a factor of 30.7 (Figure 1). But most of the ionization is above 10 km altitude and the enhancement at lower altitudes is much lower, as expected. At the 3 km level, for case 1 (minimal), we found an average increase in ionization of 38% with 79.6% enhancement at the equator and 20.4 % at the poles. For case 2 (maximal), the average enhancement was by a factor of 6, with 7.7 at the equator and 5.32 at the poles at 3 km altitude. Since higher energy primaries are unaffected by the geomagnetic field, one gets almost the same flux of particles globally. When compared to the normal GCR flux, the difference at the equator is much higher than at the poles because the normal particle flux at the equator is much lower than at the poles. For case 2, the flux of high energy primaries is even higher, which translates to an increase in the flux of high energy muons. High energy muons, because of their very small cross sections, can penetrate deeper into the atmosphere increasing ionization in the lower atmosphere. Using the most conservative estimates of low altitude cloud cover changes (10% contribution), we estimate about 4% average increase in the low altitude cloud cover from the minimal case and by 60% from the maximal case.

This large difference in our estimates is due to the large uncertainty in the HECR flux calculated from the model. For the minimal case, the change in cloud cover is probably not enough to cause a significant global temperature change. On the other hand, the cloud cover change from the maximal case is large and can certainly lead to a large drop in the global temperature and increase in precipitation. The actual cloud cover change can be anywhere between the two cases discussed here. As with earlier ozone depletion computations (Melott et al., 2010b) the estimates of effects which may impact biodiversity appear in this case to span the range from unimportant to serious.

An interesting aspect of this question concerns strontium isotope ratios. Melott and Bambach (in preparation) have noted a strong 62 My periodicity in the fluctuations of the $^{87}Sr/^{86}Sr$ ratio, with the sense that this ratio rises when biodiversity drops. Such an increase would typically be associated with uplift, but Li et al. (2007) have argued that there should be a strong temperature effect due to erosion rates of mica, raising this isotope ratio for cooler temperatures. We speculate that this effect might provide a link between the natural oscillation frequency and phase agreement of the extragalactic shock scenario and geoisotope data which is synchronous with the biodiversity signal.

More accurate computations of the atmospheric chemistry effects from cosmic rays will be possible in the future with better measurements of parameters used in the extragalactic shock model and especially with improvement in the understanding of cloud microphysics.

**ACKNOWLEDGEMENTS**

This research was supported by the NASA Program Astrobiology: Exobiology and Evolutionary Biology under grant number NNX09AM85G.

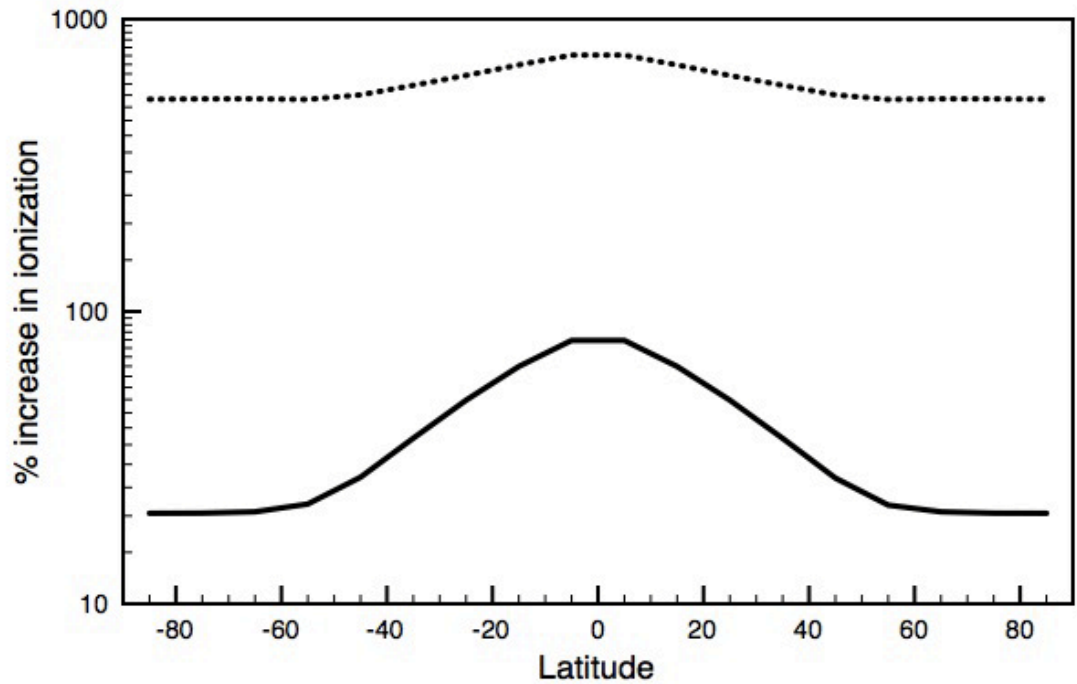

Figure 1: Percent increase in atmospheric ionization for case 1 (solid) and case 2 (dotted) at 3 km altitude. The enhancement is greater at the equator than the poles because many normal cosmic rays are guided toward the poles by the geomagnetic field, while the cosmic rays in the extra-galactic model are typically too energetic to be redirected.